\documentclass[12pt,a4paper]{article}

\RequirePackage[l2tabu, orthodox]{nag}

\usepackage{mjheppub}
\usepackage{color}
\usepackage{lipsum}
\usepackage{amsthm,amssymb,amsmath,epic,eepic,float}
\usepackage{rotating,epsfig,indentfirst,array,varioref}
\usepackage{appendix,marginnote,tikz,pgf,mathtools}
\usepackage{setspace}

\usepackage{tikz}
\usetikzlibrary{decorations.pathreplacing}
\usetikzlibrary{calc}

\usepackage{mjheppub}
\usepackage[mathscr]{eucal}
\usepackage{amsmath,amsfonts,amssymb,amsthm}
\usepackage[matrix,arrow]{xy}
\usepackage{amsthm,amsmath,latexsym,amssymb,amsfonts,amsbsy}
\usepackage{wrapfig}
\usepackage{graphicx}
\usepackage{float}

\usepackage{chngcntr}
\usepackage{amsmath}
\usepackage{mathtools}
\usepackage{amssymb}
\usepackage{braket}
\usepackage[T1]{fontenc}

\makeatletter
\def\@@bfil{\leaders \vrule \@height \ht\z@ \@depth \z@ \hfill}
\def\@bLfil{\@@bfil}
\def\@bRfil{\@@bfil}
\def\resetbraceratio{\gdef\@bLfil{\@@bfil}\gdef\@bRfil{\@@bfil}}
\def\setbraceratio#1#2{
  \let\@bLfil\relax
  \multido{\iA=1+1}{#1}{\gappto\@bLfil{\@@bfil}}
  \let\@bRfil\relax
  \multido{\iA=1+1}{#2}{\gappto\@bRfil{\@@bfil}}
}
\def\upbracefill{$\m@th\setbox\z@\hbox{$\braceld$}\bracelu\@bLfil\bracerd\braceld\@bRfil\braceru$}
\def\downbracefill{$\m@th\setbox\z@\hbox{$\braceld$}\braceld\@bLfil\braceru\bracelu\@bRfil\bracerd$}
\makeatother

\usepackage{tikz,pgf}
\usetikzlibrary{shapes}
\usetikzlibrary{calc}
\usetikzlibrary{decorations.pathmorphing}
\usetikzlibrary{decorations.pathreplacing,shapes.misc}
\usetikzlibrary{positioning}
\usetikzlibrary{arrows}
\usetikzlibrary{decorations.markings}

\def\be{\begin{equation}}
\def\ee{\end{equation}}
\def\barr{\begin{array}}
\def\earr{\end{array}}

\def\1{\tilde{1}}
\def\2{\tilde{2}}
\def\3{\tilde{3}}

\newcommand{\ba}{\begin{equation}\begin{aligned}}
\newcommand{\ea}{\end{aligned}\end{equation}}

\newcommand{\bml}{\begin{multline}}
\newcommand{\eml}{\end{multline}}

\newcommand{\CC}{\mathbb{C}}

\newcommand{\ZZ}{\mathbb{Z}}
\newcommand{\PP}{\mathbb{P}}

\newcommand{\dd}{\mathrm{d}}
\newcommand{\pd}{\partial}

\newcommand{\Res}{\mathrm{Res}}

\newcommand{\MM}{\mathscr{M}}
\newcommand{\aA}{\mathscr{A}}

\begin{document}

\title{\boldmath 
A new approach for computing  the  geometry of the moduli spaces   for  a Calabi--Yau manifold
}
\author{Konstantin Aleshkin $^{1,2}$,}
\author{Alexander Belavin $^{1,3,4}$ }

\affiliation{$^1$ L.D. Landau Institute for Theoretical Physics\\
 Akademika Semenova av. 1-A\\ Chernogolovka, 142432  Moscow region, Russia}

\affiliation{$^2$ International School of Advanced Studies (SISSA),
 via Bonomea 265, 34136 Trieste, Italy}
\affiliation{$^3$ Department of Quantum Physics\\ 
Institute for Information Transmission Problems\\
 Bolshoy Karetny per. 19, 127994 Moscow, Russia}
\affiliation{$^4$ Moscow Institute of Physics and Technology\\
Dolgoprudnyi, 141700 Moscow region, Russia}




\emailAdd{kaleshkin@sissa.it,
belavin@itp.ac.ru}

\abstract{It is known that moduli spaces of Calabi--Yau (CY) manifolds are special
  K\"ahler manifolds.
This  structure determines the corresponding low-energy effective theory which arises in 
superstring compactifications on CY manifolds. In the case, where CY manifold is given as a hypersurface
in the weighted projective space, we propose  a new procedure  for  computing the K\"ahler potential
 of the moduli space. Our method is based on   the fact  that the moduli space of CY manifolds is a marginal subspace of the Frobenius manifold which arises on the deformation space of the corresponding Landau--Ginzburg superpotential. }

\maketitle
\flushbottom

\section{ Introduction}

To compute the low-energy Lagrangian of the string theory compactified on a CY 
 manifold, one needs to know the geometry of the corresponding CY moduli space.
One way to compute this structure was proposed in \cite{COGP, BCOFHJQ}.\\
As shown in~\cite{Distances}, see also~\cite{S, CO},  the K\"ahler potential of the metric on the moduli space of CY manifold $X$
 is expressed bilinearly through periods of a holomorphic 3-form $\Omega$.
\be
\omega_{\mu} := \oint_{q_{\mu}} \Omega,
\ee
where $q_{\mu}$ form a basis of $H_3(X, \ZZ)$. 

The K\"ahler potential is computed in two steps.\\ 
The first step is to compute periods  $\omega_{\mu}$  in a special  basis of cycles $q_{\mu}$.
It was successfully done for a large number of CY manifolds and Landau--Ginzburg (LG) orbifolds in \cite{BCOFHJQ}.\\
The second step is to compute a transition matrix from $\omega_{\mu}$ to the
symplectic basis of periods $\Pi_{\mu}$.  Then K\"ahler potential is given by the bilinear form of periods  $\Pi_{\mu}$.
This second step appears to be highly non-trivial. It was done for the case of the quintic CY manifold in the
beautiful paper  \cite{COGP}.\\

In this work, we present an alternative approach to the computation of the K\"ahler potential for the case
where CY manifold is given by a hypersurface $W(x) = 0$ in a weighted projective space.
More specifically, our results immediately apply to the case when a non-degenerate CY manifold can
 be defined by a hypersurface $W_0(x)=0$ such that the function $W_0(x)$ defines an
invertible singularity~\cite{BerHub}, that is $W_0(x)$ is a polynomial with a minimal possible number
  (which is 5 for a 3-fold) of monomials 
with some further technical restriction, which is discussed in the section~\ref{sec:examples}. 

This approach is based on the connection of the CY manifold with a Frobenius manifold (FM) which arises on the monodromy invariant deformations of the singularity defined by the LG superpotential   $W_0(x) $ (see, e.g.,~\cite{Dub,Manin,hertling}). Marginal elements of the Milnor ring (quasihomogeneous deformations of the 
singularity) are in one-to-one correspondence with complex structure deformations of the CY 
manifold $X = \{W(X) = 0 \} \subset \PP^5$ and thus with $H^{2,1}(X)$. Monodromy invariant Milnor
ring is generated by these elements and is isomorphic to $H^3(X)$ where the degree of the Milnor
ring element corresponds to the Hodge decomposition. 
The moduli space of the CY manifold is a marginal subspace of this Frobenius manifold.  
This fact allows to compute easily the intersection matrix required to compute
 the K\"ahler potential.\\

We also make use of two different basises of the periods $\omega_{\mu}, \sigma_{\mu}$ of the holomorphic volume form $\Omega$
on the CY. The  basis $\omega_{\mu}$  is the same as in the papers\cite{COGP, BCOFHJQ}. 
The $\sigma_{\mu}$, in turn, is given by the connection with the chiral ring of the LG theory for the superpotential $W_0(x)$.\\  
Namely, the natural basis of the chiral ring gives a basis in the cohomology of the differential 
$D^{\pm}_{W_0} = \dd \pm \dd W_0 \wedge $. 
Periods $\sigma_{\mu}$ are then chosen as integrals over the dual basis of homology cycles 
$\Gamma^{\pm}_\mu $, which belong to the relative homology 
$H_5(\CC, \mathrm{Re}W_0(x) = \pm\infty)$. \\

The expression for the K\"ahler potential is then given  in terms of the periods $\sigma_{\mu},$ 
the holomorphic metric on the Frobenius Manifold and transition matrix $T$ 
from periods $\omega_{\mu}$ to $\sigma_{\mu}$ (see section~\ref{sec:new}):
\be
e^{-K} = \sigma^+_{\mu} \eta^{\mu\rho} M^{\nu}_{\rho} \overline{\sigma^-_{\nu}}, \quad 
\; M = T^{-1} \bar{T}.
\ee

The basis $\omega_{\mu}$ of periods has the advantage of being taken over homology
 cycles $q_{\mu}$ with integer coefficients.
Therefore, both the  K\"ahler potential and the  (holomorphic) metric of the FM are expressed in terms of the same intersection matrix of the corresponding cycles.
\ba
e^{-K(\phi)} = \omega_{\mu}(\phi) C^{\mu\nu} \overline{\omega_{\nu}(\phi)}, \\
h_{\alpha\beta} = \omega_{\alpha\mu}(0) C^{\mu\nu} \omega_{\beta\nu}(0), \\
\ea
where $(C^{-1})_{\mu\nu} = q_{\mu}\cap q_{\nu}$
and $\omega_{\alpha\mu}(\phi)$ are the different periods defined by
 integration over the same cycles.
However, it turns out that $\omega_{\alpha\mu}$ are inconvenient to use in the second equation.

On the contrary, periods of the basis $\sigma_{\mu}$ and the additional periods
 $\sigma^{\pm}_{\alpha\mu}$ defined as integrals over   homology cycles $\Gamma^{\pm}_\mu $
 with complex coefficients have more  sophisticated connection  with the K\"ahler
 potential,  but  their relation with the holomorphic metric on the FM is simple. 

We demonstrate our method on the famous Quintic hypersurface, Fermat surfaces and CY manifolds
which are defined through the invertible singularities.\\


This paper is structured as follows. In  section~\ref{sec:special}, we recall the notion of 
the Special K\"ahler geometry structure arising on the moduli spaces of CY manifolds. 
In section~\ref{sec:hyper}, we focus on the case when CY manifold is a hypersurface in a weighted projective
space and discuss periods. In  section~\ref{sec:frob} we 
review the Frobenius manifold structure connected with such CY, and we also
introduce the second set of periods. In section~\ref{sec:new}, we present our main result,
the formula for the  K\"ahler potential on the moduli space. In  section~\ref{sec:examples}, we demonstrate the efficiency of our method  with  some examples.

  \section{Special geometry}\label{sec:special}
 Here, we recall the basic facts about special  K\"ahler geometry and how it arises
 on the CY moduli space. Most of the material can be found  in \cite{S, Distances, Rolling, Candelas, CO}.\\ 
It was shown in \cite{CO} that the moduli space   $\MM$ of complex (or K\"ahler)  structures
 of a given CY manifold  is a  special  K\"ahler manifold. 
That is, if $\MM$ is  $n$-dimensional, then there are so-called special (projective) coordinates
 $z^1 \cdots z^{n+1}$ and a holomorphic homogeneous function $F(z)$ of degree 2 in $z$ called a 
prepotential such that  the K\"ahler potential $K(z)$ of the moduli space metric  is
 given by 
\be \label{specmet} e^{-K(z)} = z^a \cdot \frac{\pd \bar{F}}{ \pd \bar{z}^{\bar{a}}} - 
\bar{z}^{\bar{a}} \cdot \frac{\pd F}{ \pd z^{a}}. 
\ee
 This metric on the moduli space of complex structures  is a  metric that naturally arises from  B. deWitt metric on a space of metrics on CY manifolds.

  \paragraph{Special geometry on the complex  structure moduli space}
~\\
Consider a CY three-fold $X$. The CY moduli space of $X$ is
the space of metric perturbations of $X$ that preserve Ricci-flatness.\\
 Then, the  metric on the complex structure  CY moduli space  can be obtained\cite{CO} from natural deWitt metric  for pure  holomorhic  CY  metric deformations 
$\delta g_{\mu\nu}, \; \delta g_{\bar{\mu}\bar{\nu}}$  and given by
\be
G_{a\bar{b}} = \int_X \dd^6 y \; g^{1/2} \; g^{\mu \bar{\sigma}} g^{\nu \bar{\rho}} 
\delta_a g_{\mu\nu} \delta_{\bar{b}} g_{\bar{\sigma} \bar{\rho}} .
\ee
The CY metric deformation with two indices of the same holomorphicity  that leave the metric Ricci
 flat corresponds to elements in $H^{0,1}_{\bar{\pd}}(TX) \simeq H^{2,1}(X)$:
\be
\delta_a g_{\bar{\nu}\bar{\sigma}} \to \delta_a g_{\bar{\nu}\bar{\sigma}} g^{\bar{\sigma} \mu} 
\dd \bar{y}^{\bar{\nu}} = h^{\mu}_{a, \bar{\nu}} \dd \bar{y}^{\bar{\nu}} \sim \Omega_{\mu\nu\sigma}
h^{\sigma}_{a, \bar{\rho}} \dd y^{\mu}\wedge \dd y^{\nu}\wedge \dd y^{\bar{\rho}} \sim \chi_{a,\mu\nu\bar\rho} .
\ee
We can then rewrite the above metric  as
\be
G_{a\bar{b}} = \frac{\int_X \chi_a \wedge \bar{\chi}_{\bar{b}}}{ \int_X \Omega \wedge \bar{\Omega}}.
\ee
Here indices $a, \bar{b}$ are indices in the deformation space, that is in $H^{2,1}(X)$.
Using the following lemma (attributed to Kodaira):
\be
\pd_a \Omega = k_a \Omega + \chi_a,
\ee
we easily verify that this metric is a  K\"ahler :
\be
G_{a\bar{b}} = -\pd_a \pd_{\bar{b}} \ln \int_X \Omega \wedge \bar{\Omega} .
\ee

To obtain\cite{CO} the formulae written above  , we  choose Poincare dual symplectic basises 
 $ \alpha_a,\beta^b \in H^3(X, \ZZ)$ and $A^a,B_b\in H_3(X, \ZZ)$:
\ba
\int_{A^{a}} \alpha_{b} = \delta^{a}_{b},\quad \; \int_{A^{a}} \beta^{b} = 0, \quad 
\int_{B_{a}} \alpha_{b} = 0,\quad  \; \int_{B_{a}} \beta^{b} = \delta_{a}^{b}, \\
A^a \cap B_b = \delta^{a}_{b},\qquad \quad  \; A^a \cap A^b =0,\quad  \; B_a \cap B_b = 0.\\
\int_X \alpha_{a} \wedge \beta^{b} = \delta_{a}^{b}, \qquad  \;
\int_X \alpha_{a} \wedge \alpha_{b} =0 \;\quad 
 \int_X \beta^{a} \wedge \beta^{b} = 0.
\ea
With this, we decompose $\Omega$ as
\be \label{kahomega}
\Omega = z^a \alpha_a + F_b \beta^b,
\ee
here $z^a, F_b$ are periods of $\Omega$:
\be
z^a = \int_{A^a} \Omega, \; F_b = \int_{B_b} \Omega.
\ee
Substituting this in~\eqref{kahomega} 
we obtain

\be
e^{-K} = \int_X \Omega\wedge \bar{\Omega} = 
z^a \cdot \bar{F}_{\bar{a}} - \bar{z}^{\bar{a}} \cdot F_{ a}.
\ee
From the same  lemma we also obtain
\be
0 = \int_X \Omega \wedge \pd_a \Omega = F_a - z^b \pd_a F_b \implies F_a(z) = \frac{1}{2} \pd_a (F(z)),
\ee
where $F(z)= 1/2 z^b F_b(z)$.
Therefore, according to the defnition~\eqref{specmet} metric
 $G_{a\bar{b}} = \pd_a\bar{\pd}_{\bar{b}} \, K(z)$ is a special  K\"ahler metric 
with prepotential $F(z)$ and with special coordinates given by the period vector.\\
Using the notation for the vector of periods
\be
\Pi = \begin{pmatrix} F_{\alpha}, z^{b} \end{pmatrix},
\ee
we write the expression for the K\"ahler potential as 
\be \label{symplkahler}
e^{-K(z)} = \Pi_a \Sigma^{ab} \overline{\Pi_b},
\ee 
where $\Sigma$ is a symplectic unit, which is an inverse intersection matrix for cycles
 $A^a$ and $ B_b$.\\

Also Yukawa couplings can be expressed through the prepotential $F(z)$\cite{CO} as:
\be
\kappa_{abc} = \pd_a\pd_b\pd_c F(z) = \int_X \Omega \wedge \pd_a\pd_b\pd_c \Omega.
\ee
\paragraph{ K\"ahler  potential in an arbitrary basis of  periods.}
~\\
Using formula~\eqref{symplkahler}, we can rewrite this expression in any basis of periods 
\be
\omega_{\mu} = \int_{q_\mu} \Omega ~ ,
\ee
where $q_\mu \in H_3(X, \ZZ)$, and  then

\be\label{kahpot2}
e^{-K} = \omega_{\mu} C^{\mu\nu} \overline{\omega_{\nu}},
\ee
where $ C^{\mu\nu}$ is the inverse  marix of the intersection of cycles.

Thus, to find   the K\"ahler potential, we must compute the periods over a basis of cycles
 on CY manifold   together with their intersection matrix.

\section{Hypersurface in weighted projective space}\label{sec:hyper}

Further,  we concentrate on the case where the  CY manifold  is realized as a zero locus of
a single polynomial equation in a weighted
projective space. In this case, we establish the crucial 
relationship with the FM structure.
We consider CY 3-folds, but most of our result are generalizable to an arbitrary dimension.

\paragraph{Fundamental basis of periods}
  Let $x_1, \ldots , x_5$ be homogeneous coordinates in a weighted projective
 space and
\be
X = \{ x_1, \ldots, x_5 \; \in \PP^{4}_{(k_1, \ldots, k_5)} | \; W_0(x) = 0 \}.
\ee
For some quasi-homogeneous polynomial  $W_0(x)$, 
\be
W_0(\lambda^{k_i} x_i) = \lambda^d W_0(x_i) ,
\ee
  and 
\be
\operatorname{deg} W_0(x) = d = \sum_{i=1}^5 k_i.
\ee
The last relation ensures that $X$ is a CY manifold. 
Here $W_0(x)$ is the superpotential for the corresponding Landau--Ginzburg (LG) model,
and $W_0(x)$ defines an isolated singularity in the origin iff the CY manifold  is 
quasi-smooth~\footnote{Quasi-smoothness means that the only possible singularities of the CY are given by those of the underlying
weighted projective space}.
The moduli space of complex structures is then  given by  homogeneous polynomial
 deformations of this singularity:
\be
W(x, \phi) = W_0(x) + \sum_{s=0}^{\mu} \phi^s e_s(x) ,
\ee
where $e_s(x)$ are monomials of $x$ which have the same weight as $W_0(x)$.
In this case, the holomorphic 3-form $\Omega$ is given as a residue of a 5-form in the underlying
 affine space $\CC^5$:
\begin{multline}\label{omega}
\Omega = \frac{x_5 \dd x_1\wedge \dd x_2\wedge \dd x_{3}}{\pd W(x)/\pd x_{4}} =
 \Res_{ W(x) = 0} \frac{x_5 \dd x_1\cdots\dd x_{4}}{W(x)} = \\ =\frac{1}{2\pi i}
\oint_{|x_5|=\delta} \; \Res_{ W(x) = 0} \frac{\dd x_1\cdots\dd x_{5}}{W(x)},
\end{multline}
where the last equality is due to the homogeneity of the integrand.
Residue is understood as an integral over a small circle in the direction transversal
to the hypersurface $W(x) = 0$ at each point.

Using  explicit expression~\eqref{omega}
 for $\Omega$, we can  compute a basis of periods $\omega_{\mu}(\phi)$ as
 follows \cite{BCOFHJQ}.
We take a so-called fundamental cycle $q_1$, which is a torus in the large complex structure limit:
\be
W(x,\phi) = W_0(x) + \phi_0 \prod x_i + \sum_{s=1}^{\mu} \phi^s e_s(x), \; \phi_0 >> 1 .
\ee
In this limit, we can define an $5-$dimensional torus $Q_1 = |x_i| = \delta_i$ surrounding
the hypersurface $W(x) = 0$ in $\CC^5$. It corresponds to an $3$-dimensional torus $q_1 \subset 
X$.
Then the fundamental period is defined as an integral over this cycle
\be
\omega_1(\phi) := \int_{q_1} \Omega  = \int_{Q_1} \frac{\dd x^1\cdots\dd x^5}{W(x,\phi)} ,
\ee
and is given by a residue in its large $\phi_0$ expansion.
It was computed for many cases in the work \cite{BCOFHJQ}.\\
More periods $\omega_{\mu}$ may be obtained
as analytic continuations of $\omega_1$ in $\phi$. 
This can be conveniently done by continuing $\omega_1(\phi)$ in a small $\phi_0$ 
region using Mellin--Barnes integrals and  using the symmetry of $W_0(x)$ afterward.\\  
Namely, there is a group of \textit{phase} symmetries $\Pi_{X}$ acting diagonally on
$x_i$ and this action preserves $W_0(x)$. It doesn't preserve the deformed polynomial $W(x)$, but we
can extend this action to $\aA\; : \; G \times \{\phi^s \} \to \{\phi^s\}$ on the parameter space 
such that $W(x)$ is invariant:\\
 $$W(g\cdot x,~ \aA(g) \cdot \phi_0,~ \aA(g) \cdot \phi^s) = W(x, \phi).$$
 The moduli space  is then at most a factor of the parameter space $\{\phi^s \}/\aA$.\\
 This allows to define a set of other periods by analytic continuation,
\be
\omega_{\mu_g}(\phi) = \omega_{1}(\aA(g) \cdot \phi_0,~ \aA(g) \cdot \phi^s) \; g \in G_X .
\ee
In many cases this construction gives the whole basis of periods for the manifold $X$~\cite{BCOFHJQ}. 

\paragraph{Periods as oscillatory integrals}
~\\
The next important step  is  to transform the integrals for the periods $\int_{q_{\mu}} \Omega$ 
to the complex oscillatory form. First we have
\be
\omega_{\mu}(\phi) := \int_{q_{\mu}}\Omega = \int_{Q_{\mu}} \frac{\dd^5 x}{W(x)} ,
\ee
where $q_{\mu} \in H_{3}(X), \; Q_{\mu} \in H_5(\CC^5 \backslash W(x) = 0),$
and $Q_{\mu}$ is given by a tubular neighbourhood of $q_{\mu}$.
Now we can present them in the form
\be \label{osc}
 \int_{Q_{\mu}} \frac{\dd^5 x}{W(x)}
 = \int_{Q^{\pm}_{\mu}} e^{\mp W(x)} \dd^5 x
\ee
where
  $Q_{\mu}^{\pm} \in H_5(\CC^5, \; \mathrm{Re}W_0(x) = \pm\infty)$.  

 The map $Q_{\mu} \to Q^{\pm}_{\mu}$ is given by a contour deformation
 similarly to how it was done in \cite{AVG} and in \cite{BGK}.
This deformation can be performed due to the existence of  a natural
 isomorphism~\footnote{We are thankful to A. Givental and V. Vasiliev for
 explanation of this point.}~\cite{Vas}
\begin{equation}
H_{3}(X) \to H_5(\CC^5 \backslash W(x) = 0) =  
H_5(\CC^5, \; \mathrm{Re}W_0(x) = \pm \infty)_{w\in d\cdot\ZZ} .
\end{equation}
Here 
\be
H_5(\CC^5, \; \mathrm{Re}W_0(x)=\pm \infty)_{w\in d\cdot\ZZ}
\ee
is a subgroup of $ H_5(\CC^5, \; \mathrm{Re}W_0(x) = \pm \infty) $, 
 which we describe in the next paragraph.
Using this isomorphism, we rewrite the  periods as
\be
\omega_\mu = \omega^{\pm}_\mu = \int_{Q^{\pm}_\mu} e^{\mp W(x)} \dd^5 x ,
\ee
and from formula~\eqref{kahpot2}, we obtain
\be \label{kahlerpm}
e^{-K} = \omega^+_\mu C^{\mu\nu} \overline{\omega^-_\nu} ,
\ee
where $C^{\mu\nu}=  q_{\mu} \cap q_{\nu}=Q_{\mu}^+ \cap Q_{\nu}^-$  and
$\; Q_{\mu}^{\pm} \cap Q_{\nu}^{\pm} = 0$.
At the moment we do not know how to prove that intersections of the cycles  
$  q_{\mu} \cap q_{\nu}$  and $Q_{\mu}^+ \cap Q_{\nu}^-$ do  coincide. 
 However, this conjecture agrees with the Cecotti-Vafa formula in~\cite{CV}
for the  K\"ahler potential.  Also,   we have verified the conjecture  by direct computations in a few examples considered in section~\ref{sec:examples} and comparing our results with obtained earlier in \cite{COGP,Klemm}.\\

Taking into account that the basis  $\omega_\mu$ of periods  has  been already found for a large class of 
CY manifolds  in \cite {BCOFHJQ}, 
we need   only to know the  matrix  $C^{\mu\nu}$  to obtain the K\"ahler potential for the  Moduli space of CY manifolds in these cases.\\
To solve this problem, we will use the fact that the CY moduli space is the marginal subspace of  the Frobenius manifold \cite{Dub} which arises on the monodromy invariant 
 deformations of the singularity given by the LG superpotential  $W_0(x)$. We will do this in the next sections.

\section{ Frobenius manifold structure on the deformation space  of $W_0$} \label{sec:frob}
It was conjectured and partially demonstrated  by Gepner in \cite{Gep} that  compactifications of  superstring theory on CY manifolds  invented in \cite{CHSW} are equivalent to compactifications on $N=2$ superconformal field theories.\\
In the cases considered now  they are LG models with superpotential $W_0(x)$.
The deep connection of these models with singularity theory
was established in  \cite{LVW,Mart,VW} .
The Frobenius manifold structure appears on  the space of deformations of this singularity. 

The polynomial $W_0(x)$ in $\CC^5$, which defines a quasi-smooth
CY hypersurface in some 4-dimensional weighted projective space, is a quasi-homogeneous polynomial:
\be
 W_0(\lambda^{k_i} x) = \lambda^d W_0(x) 
\ee
with an isolated singularity in the origin.

We consider the Milnor ring of this singularity
\be
R_0 = \frac{\CC[x_1, \ldots, x_5]}{\pd_1 W_0(x) \cdot \ldots \cdot \pd_5 W_0(x)} ,
\ee
and its subring $R^Q_0$ of elements of degree divisible by degree $d$ of the singularity
~\footnote{For the quintic threefold
 $W_0(x) = \sum_i x_i^5, \; \dim \, R_0 = 1024, \; \dim R^Q_0 = 204, $ there are 101 elements of
$R^Q_0$ of the weight $d$. Q in the notation $R_0^Q$ stands for the so-called quantum
 symmetry group (see, e.g., section~\ref{sec:examples}).
Here and after one can consider the same construction with a bigger
quantum symmetries group $\ZZ_d \subset Q$.}.
 We note that this is precisely
a subring invariant under the following action of 
\be
Q = \ZZ_d \; : \; m \cdot x_j 
= e^{\frac{2\pi i m k_j }{d}} x_j,
\ee
 where $m \in Q$.
Let $e_{\mu}(x)$ be a (vector space) basis of $R^Q_0$, consisting of homogeneous monomials
 of the least  possible degree.  
There is a natural multiplication in this ring induced by multiplication of
polynomials, and there is  also a metric, turning the space of $e_{\mu}(x)$ into a 
Frobenius algebra.

The metric and struture constants are given by
\ba
\eta_{\mu\nu} = \Res \frac{e_\mu \cdot e_\nu}{\pd_1 W_0(x) \cdots \pd_5 W_0(x)}, \\
C_{\mu\nu\lambda} = C_{\mu\nu}^{\sigma} \eta_{\sigma\lambda} = \Res \frac{e_\mu \cdot e_\nu \cdot e_\lambda}{\pd_1 W_0(x) \cdots \pd_5 W_0(x)} .
\ea

We consider a general non-necessarily homogeneous deformation
\be
W(x) = W_0(x) + \sum t^{\mu} e_{\mu}(x) .
\ee
The space of parameters $t^{\mu}$ then  possesses a structure of a Frobenius manifold  $\MM_F$ . 
That is, there are a multiplication and a flat metric $h_{\mu\nu}(t)$ on
the monodromy invariant Milnor ring $R^Q$ of the deformed singularity $W(x)$:
\be
R = \frac{\CC[x_1, \ldots, x_5]}{\pd_1 W(x) \cdot \ldots \cdot \pd_5 W(x)}, \; R^Q = \{e 
\in R, \; | \; [e] \in d \cdot \ZZ\} .
\ee
Since the metric $h_{\mu\nu}$ is flat, there exist flat coorinates
 $s^{\mu}$ on $\MM_F$. The metric in these coordinates  is  the constant $\eta_{\mu\nu}$ equal  to  $h_{\mu\nu}(t=0)$. 
The structure constants of the ring $R$ are given by the third derivatives of a function $F(t)$ called the Frobenius potential,
\be\label{WDVV}
C_{\mu\nu}^{\rho}(t) h_{\rho \sigma} = \nabla_{\mu}\nabla_{\nu}\nabla_{\sigma} F(t) ,
\ee
where $ \nabla $ is Levi-Civita connection for  $h_{\mu\nu}(t)$.\\
We recall Riemannian manifold $\MM_F$ with a flat metric $h_{\mu\nu}(t)$
is called a Frobenius manifold if each tangent space of $\MM_F$ is a
Frobenius algebra with a scalar product given by $h_{\mu\nu}$ and
the structure constants satisfy  integrability condition which implies~\eqref{WDVV}
for some local function $F(t)$ (see~\cite{Dub} for more details on FM).\\

\paragraph{Frobenius manifold and Cohomology group $\bf {H^5_{D^{\pm}}(\CC^5)}$}
~\\
We consider the differentials
\be
D^{\pm} = D^{\pm}_{W_0} = \dd \pm \dd W_0\wedge .
\ee
 They define  certain cohomology groups on the $Q=\ZZ_d$-invariant differential forms in $\CC^5$.
The fifth cohomology groups $H^5_{D^{\pm}}(\CC^5)_{w \in d\cdot\ZZ}$ of this
 differentials as linear spaces  are isomorphic to the invariant Milnor ring  $R^Q$~\cite{AVG}.
This isomorphism is given by $e_{\mu}(x) \to e_{\mu}(x) \dd^5 x$ \cite{Blok,BB}. 

The cohomology group $ H^5_{D^{\mp}}(\CC^5)_{w \in d\cdot\ZZ}$ is dual to the
homology group $H_5(\CC^5, \mathrm{Re}W_0(x) = \mp \infty)_{w \in d\cdot\ZZ}$
which is by definition a $Q = \ZZ_d$ invariant subgroup of
 $H_5(\CC^5, \mathrm{Re}W_0(x) = \mp \infty)$. 
This duality follows from the self-consistency of the perfect pairing between
 these groups  defined as
\be
\langle \Gamma^{\pm}_{\mu}, \;  e_{\nu}\dd^5 x \rangle=
\int_{\Gamma^{\pm}_{\mu}} e_{\nu} \cdot e^{\mp W_0(x)} \dd^5 x .
\ee

As was already mentioned, $H_5(\CC^5, \mathrm{Re}W_0(x) = \mp \infty)_{w \in d\cdot\ZZ}$
is isomorphic to $H^3(X)$. Therefore we have an isomorphism
$R^Q \simeq H^3(X),$ moreover, under this isomorphism weight decomposition
on the one side becomes Hodge decompositon on the other. Namely there is a natural way
to construct a corresponding harmonic form from the polynomial in $R^Q$~\cite{Candelas}.

Furthermore we define a set of cycles
$\Gamma^{\pm}_{\mu}$ in the group $H_5(\CC^5, \; \mathrm{Re}W_0(x)=\pm \infty)_{w\in d\cdot\ZZ}$ by requiring that
\be \label{gammamu}
\int_{\Gamma^{\pm}_{\mu}} e_{\nu} \cdot e^{\mp W_0(x)} \dd^5 x = \delta^{\mu}_{\nu} ,
\ee
with the Kroneker matrix $\delta^{\mu}_{\nu}$.


The convenient computation technique in $H_{D^{\pm}}(\CC^5)$ introduced  in 
 \cite{BGK, BB}  can be used  to compute  the  integrals 
\be
\int_{\Gamma^{\pm}_{\mu}} e_{\nu} \cdot e^{\mp W(x,\phi)} \dd^5 x ,
\ee
$W(x, \phi) = W_0(x) + \sum_{s=0}^{\mu} \phi^s e_s(x)$. Let us review it in application to our case.

 First we expand the exponent in the
 integral above in $\phi$:
\be \label{sigma2}
\sigma^{\pm}_{\mu}(\phi) = \sum_m  \left(\prod_s\frac{(\pm\phi_s)^{m_s}}{ m_s!}\right) \int_{\Gamma^{\pm}_{\mu}} 
 \prod_s e_{s}(x)^{m_s} \, e^{\mp W_0(x)} \, \dd^5 x .
\ee
We note, that $\sigma^-_{\mu}(\phi) = (-1)^{|\mu|}\sigma^+_{\mu}(\phi),$
so we focus on $\sigma_{\mu}(\phi) := \sigma^{+}_{\mu}(\phi).$

For each of the summands in~\eqref{sigma2} form
 $\prod_s e_{s}(x)^{m_s} \, \dd^5 x$ belongs to $H^5_{D_{\pm}}(\CC^5)_{inv},$
because it is $Q-$invariant.
Thus, we can expand the form in the
 basis $e_{\mu}(x) \, \dd^5 x \in
H^5_{D_{\pm}}(\CC^5)_{inv}.$ Namely we always can find such a polynomial $4-$form $U,$ that
\be \label{sigma3}
 \prod_s e_{s}(x)^{m_s} \, \dd^5 x =
 \sum_{\nu} C_{\nu}(m) \, e_{\nu}(x) \, \dd^5 x + D_{+} U .
\ee
Therefore, the integral in~\eqref{sigma2} is reduced to
\be \label{sigma4}
\int_{\Gamma^{\pm}_{\mu}} 
 \prod_s e_{s}(x)^{m_s} \, e^{\mp W_0(x)} \, \dd^5 x = C_{\mu}(m) .
\ee

The computation of $C_{\mu}(m)$ and $U$ can be done as follows.
Let degree of $\prod_s e_{s}(x)^{m_s} \, \dd^5 x$ be $kd$. 
Since each of  differentials  $D_{\pm}$  is the sum of two quasihomogeneous terms of degree $0$ 
and degree $d$, it is convenient to present the $4-$form $U$ as a sum
 $U = \sum_{j=1}^{k-1} U_j$ with 
$\dd U_j$ quasihomogeneous of degree $j  d$ and $\dd W_0 \wedge U_j$ of degree $(j+1) d$.

Substituting to~\eqref{sigma3} arrive to the following system of equations:
\ba \label{recursion1}
 &\prod_s e_{s}(x)^{m_s} \, \dd^5 x = \dd W_0 \wedge U_{k-1}, \\
 &\dd U_{j} = \dd W_0\wedge U_{j-1}, \;   4 < j < k-1 , \\
 &\dd U_{j} = \dd W_0\wedge U_{j-1} + \sum_{\nu, \; \deg(\nu) = j d}C_{\nu}(m) \, e_{\nu}(x) \, \dd^5 x, \;   1\le j\le 4 ,
\ea
where $U_0 := 0$.

The system above can be tranformed to a recursion for the coefficients $C_{\mu}(m)$ in $m$.
 To get the recursion let us write $\dd U_{k-1} = \sum_{m'}c(m, m') \prod_s e_s(x)^{m'_s} \, \dd^5 x$ for some
constants $c(m, m').$ Then we can find these constants from  the first equation 
of~\eqref{recursion1}. This gives the relation $C_{\mu}(m) = \sum_{m'} c(m,m') C_{\mu}(m'),$ 
where $\deg \, \prod_s e_s(x)^{m'_s} < \deg \, \prod_s e_s(x)^{m_s}$.
  We will demonstrate it explicitly
 below for the quintic example.

\paragraph{ Moduli space as a subspace of the Frobenius manifold}
~\\

In general, $W(x) = W_0(x) + \sum t^{\mu} e_{\mu}(x) = 0$ does not define a
 surface in a projective space. 
This  only occurs  when $W(x)$ is quasihomogeneous, i.e. in a case of marginal deformations 
or deformations that have the same scaling property as $W_0(x)$. 
We let  denote $\{\phi^s\} \subset \{t^{\alpha}\}$ the marginal deformation  parameters.

Thus, the marginal deformations $W_0(x) + \sum \phi^{s} e_{s}(x)$ define a subspace  of
 a total  Frobenius manifold connected with the LG superpotential $ W_0$. 
This marginal subspace of the  Frobenius manifold coincides (at least locally and maybe after 
some orbifolding by a finite group) with the moduli space of the  CY manifold.

\section{Computing  the K\"ahler potential.}\label{sec:new}
\paragraph{The second basis of periods.}
~\\
In this section we use the connection of the CY moduli space to the corresponding  FM
 to find  the inverse intersection matrix of the  cycles   $C^{\mu\nu}$ ,
 $ q_{\mu} \cap q_{\nu}=Q_{\mu}^+ \cap Q_{\nu}^- $.\\

For this, we define a few  additional periods $ \omega^{\pm}_{\alpha, \mu}(\phi) $
 as integrals  of $e_{\alpha}(x) \dd^5 x$ from $H^5_{D^{\pm}}(\CC^5)$
over the cycles $Q^{\pm}_{\mu} \in H_5(\CC^5, \; \mathrm{Re}W_0(x) = \pm\infty)_{w\in d\cdot\ZZ} $ 
that have been defined earlier:
\be
\omega^{\pm}_{\alpha\mu}(\phi) = \int_{Q^{\pm}_{\mu}}  e_\alpha(x) \, e^{\mp W(x, \phi)} \dd^5 x .
\ee

In particular, the  periods $\omega^{\pm}_{1\mu}(\phi)$ coincide with the periods $\omega^{\pm}_{\mu}(\phi)$ defined above since we assume that $e_1(x) = 1$ denotes the unity in the ring $R$.

The crucial fact for the possibility to compute  the intersection matrix $C^{\mu\nu}$ is its connection  
(as proved  in the appendix)
  with the Frobenius  metric $h_{\alpha\beta}(t)$  for $t=0$ \cite{CV, Chiodo} as:\\
\be \label{hC}
\eta_{\alpha\beta} =\omega^+_{\alpha, \mu}(t=0)\; C^{\mu\nu}
\; \omega^-_{\beta, \nu}(t=0) .
 \ee
 \\
Formula (\ref{hC}) in principle allows us to determine the inverse intersection matrix
$C^{\mu\nu}$. It turns out to be convenient to use the following maneuver.\\

We define another basis of periods $\sigma^{\pm}_{\mu}(\psi)$  as integrals over the cycles $\Gamma^{\pm}_{\mu} \in H_5(\CC^5, \; \mathrm{Re}W_0(x) = \pm\infty)_{w\in d\cdot\ZZ} $:
\be \label{sigmamu}
\sigma^{\pm}_{\mu}(\psi) = \int_{\Gamma^{\pm}_{\mu}}  e^{\mp W(x,\psi)} \dd^5 x ,
\ee
where  $\Gamma^{\pm}_{\mu}$ is chosen   dual to $e_{\mu}(x)\dd^5 x$ as explained above.

Once we have an oscillatory representation for the periods $\sigma^{\pm}_{\mu}(\phi)$
 over the corresponding cycles  $\Gamma^{\pm}_{\mu}$, we can define additional integrals 
$\sigma^{\pm}_{\alpha, \mu}(\phi)$ over the same cycles as 

\ba
\sigma^{\pm}_{\alpha, \mu}(\phi) = \int_{\Gamma^{\pm}_{\mu}} e_{\alpha}(x) \, e^{\mp W(x,\phi)} \dd^5 x .
\ea
It follows from  $e_1(x) = 1$  that  $\sigma^{\pm}_{1\mu} = \sigma^{\pm}_{\mu}$.
Due to our choice of the cycles $\Gamma^{\pm}_{\mu}$ we also have 
 $\sigma^{\pm}_{\alpha, \mu}(t=0) = \delta_{\alpha, \mu}$ .

\paragraph{The construction.}
~\\
Since both $\omega^{\pm}_{\alpha, \mu}(\phi)$ and $\sigma^{\pm}_{\alpha, \nu}(\phi)$ are  basises of periods defined as the integrals over the cycles in the same space 
$H_5(\CC^5,\mathrm{Re} W_0(x) = \pm \infty)$, they are connected by some constant  matrix 
$ (T^{\pm})_\mu^{\nu}$, which is independent  of $\alpha$:
\be
\omega^{\pm}_{\alpha, \mu}(\phi) = (T^{\pm})^{\nu}_{\mu} \; \sigma^{\pm}_{\alpha, \nu}(\phi) .
\ee
Therefore,  to find the  matrix $T$, it suffices to take a few  first terms of the expansion over
 $\phi$ of the periods $\omega^{\pm}_{\mu}(\phi)$ and $\sigma^{\pm}_{\mu}(\phi)$
and to use   the equation 
\be \label{t1mu}
\omega^{\pm}_{\mu}(\phi) = (T^{\pm})^{\nu}_{\mu} \; \sigma^{\pm}_{\nu}(\phi) .
\ee

Also, knowing  that  $\sigma^{\pm}_{\alpha, \mu}(\phi=0) = \delta_{\alpha, \mu}$,
 we get 
\be \label{texpl}
\omega^{\pm}_{\alpha,\mu}(\phi=0)=(T^{\pm})^{\alpha}_{\mu} .
\ee
From eq. \eqref{hC} we then obtain
\be \label{etac}
 \eta^{\mu\nu}= (T^+)_{\rho}^{\mu} \; C^{\rho\sigma} \; (T^-)^{\nu}_{\sigma}.
\ee
From this  we find the intersection matrix $C^{\rho\sigma}$ in terms of the known Frobenius metric
$\eta^{\mu\nu}$ and the also known matrix $T$.

From~\eqref{kahlerpm},~\eqref{t1mu}, and~\eqref{etac},  we finally conclude that
\be \label{kahlerfinal}
e^{-K(\phi)} =  \sigma_\mu(\phi) \;\eta^{\mu\nu} \; M_{\nu}^\lambda \; \overline{\sigma^-_{\lambda}
(\phi)} ,
\ee
where the matrix $M^a_b = (T^{-1})^a_c \; \bar{T}^c_b$.

This formula is our main result. It gives an explicit expression for 
the K\"ahler potential $ K $ in terms of the periods $\sigma_\mu(\psi)$,  \\
FM metric $\eta_{\mu\nu}$  and matrix $M^\mu _\nu$.
All these data can be computed exactly as it has been explained above.\\

It makes sense to stress that having the exact expression for $\omega^{\pm}_{\nu}(\phi)$,
 we  obtain  the exact and explicit expressions for the periods $\sigma^{\pm}_{\mu}(\phi)$ from~\eqref{t1mu}:
\be
\sigma^{\pm}_{\mu}(\phi) = \left((T^{\pm})^{-1}\right)^{\nu}_{\mu} \; \omega^{\pm}_{\nu}(\phi) .
\ee
In terms of  the periods  $\sigma^{\pm}_{\mu}(\phi)$  expression~\eqref{kahlerfinal}
for the K\"ahler potential has a more convenient form for calculating the metric on the CY moduli space.

\section{Examples} \label{sec:examples}

\paragraph{Quintic CY threefold }
~\\
This CY manifold is  defined as
\begin{equation}
X_{\psi} = \{x_i \in \PP^4 \; | \; W_{\psi}(x) = 
x_1^5+x_2^5+x_3^5+x_4^5+x_5^5
 -5\psi x_1x_2x_3x_4x_5 = 0 \} .
\end{equation}
Here we set $\phi_0 = 5\psi$ to connect with~\cite{COGP}.\\
We compute the complex structure moduli space metric for this one-parameter family.
In this case, the phase symmetry(see section~\ref{sec:hyper}) is $\ZZ_5^5$ and the induced action $\aA$ on
the one-dimensional space $\{\psi \}$ is $\ZZ_5: \psi \to e^{2\pi i/5} \psi$.\\
 This space is the whole complex structure  moduli space of the quotient $X / \ZZ^3_5 =: \hat{X}$,
 that is the mirror manifold of the original quintic. In particular, $h_{1,1}(\hat{X}) = 101, \; h_{2,1}(\hat{X}) = 1$.\\
 We choose cycles $\Gamma^{\pm}_{\mu}$ dual to
the cohomology classes $\dd^5 x, \; \prod x_i \cdot \dd^5 x, \; \prod x^2_i \cdot \dd^5 x,
 \; \prod x^3_i \cdot \dd^5 x$, that form a basis  in the cohomology subgroup corresponding to  the subring of  Milnor ring elements of weight $\in 5\ZZ$ and invariant under the  $\ZZ^3_5$ action of the phase symmetry.
For the periods, the algorithm, described above gives:
\be
\label{sigma2q}
\sigma^{-}_{\mu}(\psi) = \sum_m  \left(\frac{(- 5\psi)^{m}}{ m!}\right) \int_{\Gamma^{-}_{\mu}} 
 (x_1x_2x_3x_4x_5)^{m} \, e^{ \sum_i x_i^5} \, \dd^5 x .
\ee
Let $m = 5 n + \nu, \; \nu < 5$. We want to expand 
\be \label{quintrec1}
\prod x_i^{5 n + \nu} \, \dd^5 x =
 \sum_{\nu} C_{\mu}(m) \, (x_1x_2x_3x_4x_5)^{\mu} \, \dd^5 x + D_{+} U .
\ee
Note that 
\begin{multline} \label{rec}
D_+ \left(\frac{1}{5}x_1^{5n + k-4} \, f(x_2, \cdots, x_5) \, \dd x_2 \wedge \cdots \wedge \dd x_5
\right) = \\ =
\left[x_1^{5n+k} + \left(n+\frac{k-4}{5}\right) x_1^{5(n-1) + k} \right]  \, f(x_2, \cdots, 
x_5) \, \dd^5 x ,
\end{multline}
where $f$ denotes an arbitrary polynomial. Applying it to $f = (x_2x_3x_4x_5)^{5n+k}$ we get
\be
\prod x_i^{5 n + \nu} \, \dd^5 x = -\left(n+\frac{k-4}{5}\right) \,x_1^{5(n-1)+\nu}
(x_2x_3x_4x_5)^{5 n + \nu} \, \dd^5 x \text{  modulo  }D_+(\cdots) 
\ee
We use an analogous formula for permuted $x_1, \ldots, \; x_5$ and appropriate $f$ to get the following recursion:
\be
\prod x_i^{5 n + \nu} \, \dd^5 x = -\left(n+\frac{k-4}{5}\right)^5 \,\prod x_i^{5 (n-1) + \nu} \, \dd^5 x
 + D_+ \tilde U ,
\ee
which implies 
$$C_{\mu}(m) = -\left(n+\frac{k-4}{5}\right)^5 C_{\mu}(m-5).$$
 This recursion is
 immediately solved
\be
\prod x_i^{5 n + \nu} \, \dd^5 x = (-1)^{n}\frac{\Gamma(n+\nu/5)}{\Gamma(\nu/5)}^5 \,\prod x_i^{\nu} \, \dd^5 x
 + D_+ U ,
\ee
$$C_{\mu}(m) = \delta_{\mu,\nu} (-1)^n\frac{\Gamma(n+\nu/5)}{\Gamma(\nu/5)}^5 .$$

Plugging it into~\eqref{sigma2q} we obtain
\begin{equation}
\sigma^{\pm}_{{\mu}}(\psi) = \frac{(\pm 1)^{{\mu}-1}}{\Gamma(\mu/5)^5 5^{\mu}\psi}\sum_{n=0}^{\infty}
 \frac{\Gamma^5(n+\mu/5)} {\Gamma(5n+\mu)}(5\psi)^{5n+\mu} ,
\end{equation}
which have the following behaviour $(\pm \psi)^{\mu-1}/\Gamma(\mu) + O(\psi^{\mu+4})$.

The fundamental period for the quintic is defined as a residue of a holomorphic three-form
\be
 \frac{x_5 \dd x_1\wedge\dd x_2\wedge\dd x_3}{\pd P_{\psi}/\pd x_4} ,
\ee
that is given by an integral over a cycle $q_1$,  which is  three-dimensional torus~\cite{COGP}.
Its analytic continuations as explained  above give the whole basis of periods
 in a basis of cycles with integral
coefficients:
\be
\omega_{\mu}(\psi) = 
\sum_{m=1}^{\infty} 
\frac{e^{4\pi i m/5}\Gamma(m/5)(5e^{2\pi i (\mu-1)/5}\psi)^{m-1}}{\Gamma(m)\Gamma^4(1-m/5)} ,
\quad \; |\psi| < 1 ,
\ee
in this case $\omega_{\alpha,\mu}(0)$ equals to $\pd_{\psi}^{\alpha-1}\omega_{\mu}(\psi)$
for $\alpha = 1,..,4$.

Therefore, taking the first four terms of the expansion of the periods above and using~\eqref{texpl}
we obtain
\be
T^{\mu}_{\nu} = \frac{e^{2\pi i (\nu-1) (\mu-1)/5}e^{4\pi i \nu/5}5^{\nu-1}\Gamma( \nu/5) }
{\Gamma^4(1-\nu/5)} .
\ee
The FM holomorphic metric in this case
\be
\eta = \mathrm{antidiag}(1,1,1,1) .
\ee
Finally, we obtain $\eta \, M = \eta \, T^{-1}\bar{T}$ and the K\"ahler potential for the metric
  according to our main formula~\eqref{kahlerfinal}:

\begin{equation}
e^{-K(\psi)} = \frac{\Gamma^5(1/5)}{125\Gamma^5(4/5)} \sigma^{+}_{11}\overline{\sigma^{-}_{11}}
 +\frac{\Gamma^5(2/5)}{5\Gamma^5(3/5)} \sigma^{+}_{12}\overline{\sigma^{-}_{12}} 
 +\frac{5\Gamma^5(3/5)}{\Gamma^5(2/5)}\sigma^{+}_{13}\overline{\sigma^{-}_{13}}  +\frac{125\Gamma^5(4/5)}{\Gamma^5(1/5)}
 \sigma^{+}_{14}\overline{\sigma^{-}_{14}} .
\end{equation}
In particular, 
\be
G_{\psi\bar{\psi}}(0) = 25\frac{\Gamma^5(4/5)\Gamma^5(2/5)}{\Gamma^5(1/5)\Gamma^5(3/5)} .
\ee

\paragraph{Fermat hypersurfaces}
~\\
The direct generalization of the quintic is a Fermat hypersurface, which  is the one given by the equation
\be
W_0(x) = \sum_{i=1}^5  x_i^{n_i}, \qquad \; n_i = d/k_i, \quad \; \sum k_i = d ,
\ee
and  the degree $d$ is equal to the least common multiple of $\{ k_i \}  $.\\
 As in the case above, we consider a one-dimensional deformation $W(x, \phi_0) = W_0(x) + 
\phi_0 \prod_{i=1}^5 x_i$.
The phase symmetry group is  $\Pi_X  = \ZZ_{n_1}\times \cdots \times \ZZ_{n_5}$.
The lifted action on $\phi_0$ is $\ZZ_d: \; \phi_0 \to \zeta \phi_0, \; \zeta = e^{2\pi i/d}$.
We take the expression for the fundamental period from~\cite{BCOFHJQ}:
\be
\omega_1(\phi_0) 
= \sum_{r=1}^{d-1} A(r) \frac{\phi_0^{r-1}}{\Gamma(r)} + O(\phi_0^{d - 1}) .
\ee
and 
\be
A(r)= \frac{ (-1)^{r-1}e^{\frac{-\pi i r}{d}}}
{\sin \frac{r\pi}{d} \prod_{i=1}^5 \Gamma(1 - \frac{k_j r}{d})} .
\ee

We note that $A(r)$ vanishes if $k_i r/d \in \ZZ, $ i.e.  $r/n_i \in \ZZ$.
According to the general  analytic continuation procedure  
\be
\omega_{\mu}(\phi_0) = \sum \zeta^{(r-1)(\mu-1)} A(r) \frac{\phi_0^{r-1}}{\Gamma(r)}
+ O(\phi_0^{d - 1}) .
\ee
Using the definitions ~\eqref{gammamu} and~\eqref{sigmamu} we obtain
\be
\sigma^+_{\mu}(\phi_0) = \frac{\phi_0^{\mu-1}}{\Gamma(\mu)} + O(\phi_0^{\mu + d-2}), \quad
\; \mu/n_i \notin \ZZ, \; \text{otherwise 0} .
\ee
This latter condition implies that $\omega_{\mu}$ form a basis in the periods of $\Omega$ deformed by 
$\phi_0$. 
We obtain the transition matrix 
\ba
&T^{\mu}_{\nu} = \zeta^{(\mu-1)(\nu-1)}A(\mu),\quad \; \mu/n_i \notin \ZZ, \quad \; \nu/n_i \notin \ZZ \\
&(T^{-1})^{\lambda}_{\mu} = \frac{\bar{\zeta}^{(\lambda-1)(\nu-1)}}{\tilde{d}-1} \frac{1}{A(\mu)} ,
\ea
and the real structure
\be
M^{\mu}_{\nu} = \frac{\bar{A}(\mu)}{A(d - \mu)} \delta_{\mu+\nu, d}~.
\ee
In this case, $\eta_{\mu, \nu} = \delta_{\mu+\nu, d}$ and therefore
\be
e^{-K(\phi_0)} = \sum_{\mu = 1, \; \mu/n_i \notin \ZZ}^{d-1} \prod_{i=1}^5 
\gamma\left( \frac{k_i \mu}{d} \right) \sigma^+_{\mu}(\phi_0) \overline{\sigma^-_{\mu}(\phi_0)} ,
\ee
where $\gamma(x) =\Gamma(x)/\Gamma(1-x)$ and
\be
\sigma^{\pm}_{\mu}(\phi_0) = \pm \sum_{R=0}^{\infty}  
\frac{\phi_0^{\mu-1+d R}}{ \Gamma(dR + \mu)}
\prod_{j=1}^5 \frac{\Gamma\left(k_j(R + \frac{\mu}{d})\right)}{\Gamma(\frac{k_j \mu}{d})} .
\ee
From this we get a formula for the metric itself
\be
G_{\phi_0\overline{\phi_0}} = \prod_{i=1}^5 \left(\gamma\left(\frac{k_i\mu_0}{d}\right)
\gamma\left(1-\frac{k_i}{d}\right)\right) \frac{|\phi_0|^{2(\mu_0-1)}}{\Gamma(\mu_0)^2} + 
O(|\phi_0|^{2\mu_0}),
\ee
where $\mu_0$ is the least integer $1\le\mu_0<d$ such that $(\mu_0+1)/n_j \ne \ZZ$.
The last formula reproduces the  known results 
for CY manifolds $\PP^4_{(2, 1, 1, 1, 1)}[6]$, $\PP^4_{(4, 1, 1, 1, 1)}[8]$ and
 $\PP^4_{(5, 2, 1, 1, 1)}[10]$ obtained in \cite{Klemm}.

\paragraph{The case  of invertible polynomial}
~\\
We assume that the above approach  is applicable to the case of CY manifold defined in terms of the hypersurface in weighted projective spaces of the type described in~\cite{BCOFHJQ}, i.e.,~the hypersurfaces
 whose defining polynomial is
\be \label{5mon}
W_0(x) = \sum_{j=1}^5 \prod_{i=1}^5 x_i^{a_{ij}}, \qquad \; \sum k_i a_{ij} = d , 
\ee
and
$$ \sum k_i = d.$$
In this case periods are given in terms of the \textit{mirror} CY manifold $\hat{X}$.
We briefly describe the construction of $\hat{X}$ following~\cite{BerHub}.\\
The polynimial $W_0(x)$ has a group $\Pi_X$ of phase symmetries represented as
\be
\Pi_X = Q_X \times G_X ,
\ee
where $Q_X$ is a \textit{quantum symmetry} group 
$\simeq (\ZZ_d: k_1, \cdots, k_5)$~\footnote{Notations are the same as in~\cite{BerHub},
that is $1 \in (\ZZ^r : a_1, \cdots, a_5)$ acts as $x_i \to e^{2\pi i a_i/r} x_i$. We note that action
of the quantum symmetries on $X$ is trivial.}.
The complement to $Q_X$ in $\Pi_X$ is called a \textit{geometric symmetry} group $G_X$.
For mirror manifolds the total phase symmetry is unchanged whereas roles of quantum and
geometric symmetries switch:
\be \label{GQ}
G_X = Q_{\hat{X}}, \qquad \; Q_X = G_{\hat{X}} .
\ee
To build such a mirror, we must  first to consider a polynomial $\hat{W}_0(x)$ with a transposed
matrix of exponents $\hat{a}_{ij} = a_{ji}$,
\be
\hat{W}_0(x) = \sum_{j=1}^5 \prod_{i=1}^5 x_i^{\hat{a}_{ij}}, \qquad  \; \sum \hat{k}_i a_{ji} =\hat{d} ,  
\ee
and 
$$\sum \hat{k}_i = \hat{d}.$$ \\
Here $\hat{k_i}$ and $\hat{d}$ are uniquely defined by the reqirement that the equalities
 above are satisfied.\\
 This polynomial has the same group of phase symmetries, however generically the condition~\eqref{GQ}
is not fulfiled, i.e. its quantum symmetry is smaller,
 then geometric symmetry of the original hypersurface.\\ 
To get a mirror we need to enlarge quantum symmetry of $\{\hat{W}_0(x) = 0 \}$.
For this purpose we take a quotient of the hypersurface $\{ \hat{W}_0(x) = 0 \} / H$, where
$H$ is some subgroup of phase symmetries which is to be found in each case~\cite{BerHub}. 

Thus, computing complex moduli space for the manifold $X$ (or $\hat{X}$) we compute also
a complexified K\"ahler moduli space metric for the mirror CY through the mirror map.
 
The periods $\omega_{\mu}(\phi)$ in this case were computed in~\eqref{5mon} in \cite{BCOFHJQ} and,
if we set all parameters  $\phi^s $ (but not $\phi_0$) equal to zero for simplicity, then we have:
\ba
\omega_{1}(\phi_0)  = \sum_{r=1}^{\hat{d}-1} A(r) \frac{\phi_0^{r-1}}{\Gamma(r)} + 
O(\phi_0^{\hat{d}-1})\\
A(\mu) = (-1)^{\mu} \frac{\pi}{\hat{d}\sin \frac{\pi\mu}{\hat{d}}} 
\prod_{j=1}^5 \frac{1}{\Gamma(1-\frac{\hat{k}_j\mu}{\hat{d}})} .
\ea
For our general method to work, this must give all relevant periods. Basically we must 
 check that all possible periods are obtained from this one (with all $\phi^s \ne 0$)
 by phase-symmetry analytic continuations.  In other words it is necessary to verify the relation
\be
\dim \langle \omega_{0} (\aA(g) \cdot \phi) \rangle_{g \in G_X} = \dim H_3(X) .
\ee
This was certainly the case in the preceding examples, but we are not
aware of this fact in general (it is so in all  examples).
As in the previous example, in the one-modulus case we obtain
\be
e^{-K(\phi_0)} = \sum_{\mu=1, \; \mu \hat{k}_i/\hat{d} \notin \ZZ}^{\hat{d}-1} 
 \eta^{\mu, \hat{d}-\nu} \prod_{j=1}^5 
\gamma\left(\frac{\hat{k}_j \mu}{\hat{d}}\right)
\sigma^+_{\mu}(\phi_0) \overline{\sigma^-_{\nu}(\phi_0)} .
\ee
For this formula to hold the number of linearly independent elements 
$\prod_{i=1}^5 x^n_i \dd^5 x \in H^5_{D^{\pm}}(\CC^5)$ should be equal to the number of $1 \le \mu 
< \hat{d}, \; \mu k_i/d \ne \ZZ$.\\

If $\hat{H} \subset G_X$, we may also restrict all the computations to the case of 
$\hat{H}-$invariant deformations and homology elements respectively (to describe a mirror
 manifold, for instance). 

\section{Conclusion}

A new  method for computing the CY moduli space metric is proposed. It can be used
for a larger class of of CY manifolds than this have been done before. This method does not demand using of
Picard--Fuchs equations. Instead of this, the cohomology technique  used in \cite{BB} for computing of the FM flat coordinates~\cite{BDM, BU} can be applied for computing the periods. Therefore, as we expect, it can be applied for 
the computations of the CY moduli space geometry in cases when the dimesion of the moduli space 
is more than one.\\
The FM structure naturally arising from an N=2 superconformal theory plays a 
significant role in our computation. The result is given in terms of the topological metric
on the latter manifold and two basises of periods, both of which we are able to compute 
avoiding the complicated direct computation of the symplectic basis
 of periods.\\
Our methods  have been applied in this paper to CY manifolds, given by one polynomial equation,
 in particular to the case of Fermat hypersurfaces and to the famous quintic threefold.  We suppose that the same approach  can be applied  to CY manifolds of a more general type. We also work on
the problem of computing the moduli space metric around other interesting points in the moduli
 spaces, such as the large complex structure limit.

\section{ Acknowledgements}
We are grateful to M. Bershtein, B. Everett, D. Gepner, A. Givental, A. Okounkov, A. Rosly,
 A. Varchenko and V. Vasiliev
  for the  useful discussions. We are also thankful to T. H\"ubsch for his useful comments during
the preparation of the paper. The work of A. B. on the main results presented in sections 1--5 was
 performed at IITP with the financial support of the Russian Science Foundation (Grant No.14-60-00150).
 The work of K.A. was supported by the Foundation
for the advancement of theoretical physics ``BASIS''.
\bigskip

\appendix{\bf APPENDIX }
\counterwithout{equation}{section}
\paragraph {The proof of the relation~\eqref{hC}}
~\\
Following \cite{Chiodo} here we want to prove the first equality in~\eqref{hC}, that
is
\begin{equation} \label{hh}
h_{ab} = \Res \frac{e_a \cdot e_b \; \dd^n x}{\pd_1 W_0 \cdots \pd_n W_0}
=\int_{Q^+_{\mu}} e_a \; e^{-W_0} \dd^n x \; C^{\mu\nu} \int_{Q^-_{\nu}} e_b \; e^{W_0} \dd^n x .
\end{equation}
To do this consider a small Morse function perturbation $W(x, t) = W_0(x) + e_{a} t_a$ so that
$0$ - critical point of $W$ becomes a set of Morse points $p_1, \ldots, p_{\mu}$ and consider a
 bilinear form
\be \label{hhz}
h_{ab}(t,z) = \int_{Q^+_{\mu}} e_a \; e^{-W(x,t)/z} \dd^n x \; C^{\mu\nu}
 \int_{Q^-_{\nu}} e_b \; e^{W(x,t)/z} \dd^n x .
\ee
First of all we notice, that
\be
h_{ab}(t=0, z) = z^k \cdot h_{ab}(t=0, 1),
\ee
because if $t=0,$ we can absorb $z$ by coordinate transform $x_i \to z^{k_i/d} x_i$.\\

 Then we choose
 basis of cycles of so-called Lefschetz thimbles: $L^{\pm}_i$ starts from $p_i$ and goes along
the gradient of $\mathrm{Re}(W(x,t))$ in positive/negative direction. With certain orientation their
intersection numbers are $L^+_{\mu} \cap L^-_{\nu} = \delta_{\mu\nu}$.

Then rhs of~\eqref{hhz} becomes in this basis:
\be\label{hh1}
\sum_{i=1}^{\mu} \int_{L^+_{i}} e_a \; e^{-W(x,t)/z} \dd^n x \;
 \int_{L^-_{i}} e_b \; e^{W(x,t)/z} \dd^n x .
\ee
Using stationary phase expansion as $z\to 0$ we obtain for a period:
\be
\int_{Q^+_{i}} e_a(x) \; e^{-W(x,t)/z} \dd^n x = \pm\frac{(2\pi z)^{N/2}}{\sqrt{\mathrm{Hess} W(p_i,t)}}(e_a(p_i) + O(z)) .
\ee
Inserting in the formula~\eqref{hh1} we get
\begin{multline}
h_{ab}(t,z) = \pm\sum_{i=1}^{\mu} (2\pi i z)^N \frac{e_a(p_i) \cdot e_b(p_i)}{\mathrm{Hess}(W(p_i,t))}(1 + O(z)) = \\
= (2\pi i z)^N \left(\Res \frac{e_a \cdot e_b \dd^n x}{\pd_1 W \cdots \pd_N W} + O(z)\right) .
\end{multline}
By analytic continuation it holds for $t=0$. Now we recall that $h_{ab}(0, z) = z^k \cdot h_{ab}(0,1)$.
 The required equality now follows from the previous formula.



\begin{thebibliography}{10}

\bibitem{COGP}
P.~Candelas, X.~de la Ossa,P. S. ~Green and L.~ Parkes,
{\it {A Pair of Calabi--Yau manifolds as an exactly soluble superconformal
      field theory}}, {\em Nucl.Phys.}  {\bf B359} (1991) 21--74


\bibitem{BCOFHJQ}
P.~Berglund, P.~Candelas, X.~de la Ossa, A.~Font, T.~Huebsch, D.~Jancic and F.~Quevedo,
{\it { Periods for Calabi--Yau and Landau--Ginzburg vacua }},
{\em Nucl.Phys.}  {\bf B355} (1991) 455


\bibitem{Distances}
P.~Candelas, P.~Green, T.~H\"ubsch,
{\it { Finite Distances Between Distinct Calabi-Yau Manifolds }},
{\em Phys.Rev.Lett.} {\bf 62} (1989) 1956-1959

\bibitem{S}
A.~Strominger,
{\it {Special   geometry, }}
{\em  Commun. Math. Phys.} {\bf 133} (1990) 163

\bibitem{CO}
P.~Candelas, X.~de la Ossa,
{\it {Moduli  Space of Calabi--Yau  manifolds}},
{\em Nucl.Phys.}  {\bf B335} (1991) 355-481

\bibitem{BerHub}
P.~Berglund and T.~H\"ubsch
{\it{A generalized constraction of Mirror manifolds}},
{\em Nucl.Phys.}  {\bf B393} (1993) 377


\bibitem{Dub}
B.~Dubrovin, 
{\it {Integrable systems in topological field theory}},
{\em Nucl.Phys.} {\bf B379} (1992) 627--689


\bibitem{hertling}
Claus Hertling.
\newblock {\em Frobenius Manifolds and Moduli Spaces for Singularities}.
\newblock Cambridge Tracts in Mathematics. Cambridge University Press, 2002.

\bibitem{Manin}
Yu~I Manin.
\newblock {\em Frobenius manifolds, quantum cohomology, and moduli spaces}.
\newblock 2007.


\bibitem{Rolling}
P.~Candelas, P.~S.~Green, T.~H\"ubsch,
{\it { Rolling Among Calabi-Yau Vacua }},
{\em Nucl.Phys.} {\bf B330} (1990) 49-102


\bibitem{Candelas}
P.~Candelas,
{\it { Yukawa couplings between (2,1)-forms }},
{\em Nucl.Phys.}  {\bf B298} (1988) 458-492

\bibitem{AVG}
A.~Arnold, A.~Varchenko, S.~ Gusein-Zade, 
{\it {Singularities of Differentiable Maps}},
{\em Birkhauser} 1985


\bibitem{BGK}
A.~Belavin, D.~Gepner, Ya.~Kononov,
{\it {Flat coordinates for Saito Frobenius manifolds and String theory }},
{\em arXiv}:1510.06970

\bibitem{Vas}
V.~Vasiliev's  Letter to A.~Belavin

\bibitem{CV}
S.~Cecotti and C.~Vafa,
{\it {Topological--anti--topological fusion}},
{\em Nucl.Phys.} {\bf B367} (1991) 359--461

\bibitem{Klemm}
A.~Klemm, S.~Theisen,
{\it{Recent efforts in the computation of string couplings}},
{\em TMF} (1993)  95, 2


\bibitem{Gep}
D.~Gepner,
 {\it Exactly solvable string compactifications on manifolds of SU(N) Holonomy},
{\it Phys. Lett.} {\bf 199} (1987) 380-388.

\bibitem{CHSW}
P.~Candelas, G.~Horowitz, A.~Strominger and E.~Witten
{\it{Vacuum configurations for Superstrings}},
{\em Nucl.Phys.} {\bf B258} (1985) 46--74

\bibitem{LVW}
W.~Lerche,C.~Vafa and N.~Warner,
{\it{Chiral rings in N = 2 superconformal theories}},
{\em Nucl.Phys.} {\bf B324} (1989) 327--474

\bibitem{Mart}
E.~Martinec, {\it Algebraic geometry and effective lagrangians}.
{\em Phys. lett.} {\bf 217B}(1989) 431.

\bibitem {VW}
C.~Vafa, N.~Warner{\it }
{\em Phys.Lett.} {\bf 218B} (1989) 51.

\bibitem{Blok}
B.~Blok, A.~Varchenko, {\it Topological conformal field theories and the flat coordinates},
{\em Internat. J. Mod. Phys.}{\bf 7} (1992) 1467-1490.

\bibitem{BB}
A.~Belavin, V.~Belavin,
 {\it{Flat structures on the deformations of Gepner chiral rings}},
{\em J. High Energy Phys.}, 10, 128 (2016)

\bibitem{BDM}
A.~Belavin, B.~Dubrovin, B.~Mukhametzhanov,
 {\it{Minimal Liouville gravity correlation numbers from Douglas string equation}},
{\em J. High Energy Phys.}, 01, 156 (2014)

\bibitem{BU}
V.~Belavin,
 {\it{Unitary Minimal Liouville Gravity and Frobenius manifolds}},
{\em J. High Energy Phys.}, 07, 129 (2014)


\bibitem{Chiodo}
A.~Chiodo, H.~Iritani and Y.~Ruan,
{\it{Landau--Ginzburg/Calabi--Yau correspondence, global mirror symmetry and Orlov equivalence}},
{\em arXiv}:1201.0813 [math.AG]




\end{thebibliography}
\end{document}